\documentclass[12pt]{article}
\usepackage{graphicx,wrapfig,lipsum}

\newcommand\pubnumber{SNSN}
\newcommand\pubdate{\today}

\def\saha{Saha Institute of Nuclear Physics, Kolkata, India}
\def\gsi{GSI Helmholtzzentrum fur Schwerionenforschung GmbH, Darmstadt}
\def\spain{Universidade de Santiago de Compostela,Santiago de Compostela, Spain}
\def\TU{Technische Universitat Darmstadt, Darmstadt, Germany}
\def\Brazil{Instituto Tecnologico de Aeronautica, Sao Jose dos Campus, Brazil}

\def\liverpool{University of Liverpool, Liverpool, United Kingdom}
\def\infn{INFN, LNL, Legnaro, Italy}
\def\spain{Universidad Complutense de Madrid, Avenida Complutense, Madrid, Spain}

\def\chalmer{Fundamental Fysik, Chalmers Tekniska Hogskola,  Goteborg, Sweden}
\def\kvi{KVI-CART,  The Groningen university, The Netherlands}
\def\mun{ Technische Universit\"at M\"unchen,  Garching, Germany}

\def\fz{Helmholtz-Zentrum Dresden-Rossendorf, Dresden, Germany}%
 
\def\support{\footnote{ushasi.dattapramanik@saha.ac.in, Alexander Von Humboldt,Germany,  DAE, Govt. of India}}

\def\Title#1{\begin{center} {\Large #1 } \end{center}}
\def\Author#1{\begin{center}{ \sc #1} \end{center}}
\def\Address#1{\begin{center}{ \it #1} \end{center}}

\newcommand\pubblock{\rightline{\begin{tabular}{l} \pubnumber\\
         \pubdate  \end{tabular}}}
\newenvironment{Abstract}{\begin{quotation}  }{\end{quotation}}
\newenvironment{Presented}{\begin{quotation} \begin{center} 
             PRESENTED AT\end{center}\bigskip 
      \begin{center}\begin{large}}{\end{large}\end{center} \end{quotation}}

\def\beq{\begin{equation}}
\def\eeq#1{\label{#1}\end{equation}}
\def\eeqn{\end{equation}}

\def\beqa{\begin{eqnarray}}
\def\eeqa#1{\label{#1}\end{eqnarray}}
\def\eeqan{\end{eqnarray}}

\let\bar=\overbar

\def\Dslash{\not{\hbox{\kern-4pt $D$}}}
\def\dslash{\not{\hbox{\kern-2pt $\del$}}}

\def\msb{{\bar{\ssstyle M \kern -1pt S}}}

\begin{document}
\begin{titlepage}
\pubblock

\vfill
\Title{Weakly Bound Neutron-Rich Nuclei and Cosmic Phenomena}
\vfill
\Author{ Ushasi Datta \support, A.Rahaman, S.Chakraborty, B.K.Agrawal}
\Address{\saha}
\Author{T.Aumann, K.Boretzky, C.Caesar, H.Emling, H.Geissel, C.Langer, T.Le Bleis, 
Y.Leifels, J.Marganiec, G.M\"unzenberg, C.Nociforo, R.Plag, V.Panin,
R.Reifarth,  M.V.Ricciardi, D.Rossi, C.Scheidenberger, H.Simon, S.Typel,  V.Volkov, F.Wamers, J.S.Winfield, H.Weick  }
\Address{\gsi}
\Author{B. Jonson, H.Johansson, T.Nilsson}
\Address{\chalmer}
\Author{A. Wagner}
\Address{\fz}
\Author{G.De.Angelis}
\Address{\infn}
\Author{B.V.Carlson}
\Address{\Brazil}
\Author{N. Kalantar-Nayestanaki,M.A. Najafi,C. Rigollet}
\Address{\kvi}
\Author{T. Kr\"oll, H.Scheit}
\Address{\TU}
\Author{ R. Kr\"ucken}
\Address{\mun}
\Author{ S.Beceiro-Novo, D.Cortina, P.Diaz Fernandez}
\Address{\spain}
\Author{J.T.Taylor, M.Chartier}
\Address{\liverpool}

\begin{Abstract}
The  single particle and bulk properties  of the  neutron-rich nuclei 
 constrain fundamental issues in nuclear physics and nuclear astrophysics like the 
limits of existence of quantum many body systems
(atomic nuclei), the equation of state of neutron-rich matter,  neutron star, nucleosynthesis, 
evolution of stars, neutron star merging
 etc.. The state of the art of  Coulomb breakup of the neutron-rich nuclei  has been used to explore those
properties. Unambiguous information on detailed 
components of the ground-state wave-function along with quantum numbers  of the valence 
neutron of the nuclei  have been obtained from the measurement of threshold strength along
 with the $\gamma$-rays spectra of the core following Coulomb breakup. The shape of this threshold strength 
is a finger-print of the quantum numbers of the nucleon. We investigated the ground-state
 properties of the neutron-rich Na, Mg, Al nuclei around N $\sim$ 20 using this method at GSI, Darmstadt.
 Very clear evidence has been observed for melting and merging of long cherished magic
 shell gaps at N = 20, 28. The evanescent neutron-rich nuclei  imprint their existence in 
stellar explosive scenarios (r-process etc.). Coulomb dissociation (CD) is one of the important indirect measurements 
of the capture cross-section  which may provide valuable input to the model for star evolution process, 
particularly the r-process.
 Some valuable bulk properties of the neutron-rich nuclei like the density dependent symmetry energy, 
 neutron skin etc. play a key role in understanding cosmic phenomena and these properties have been studied
via electromagnetic excitation.  Preliminary results of electromagnetic excitation
 of  the neutron-rich nucleus, $^{32}$Mg are presented.
\end{Abstract}
\vfill
\begin{Presented}
 Thirteenth Conference on the Intersections of Particle and Nuclear Physics (CIPANP 2018)\\
\end{Presented}
\vfill
\def\thefootnote{\fnsymbol{footnote}}
\setcounter{footnote}{0}
\end{titlepage}
  
  
\section{Introduction}
Recent advancement of accelerator physics along with detector technologies  provide a unique
laboratory of neutron-rich nuclei where one can explore various fundamental issues 
in nuclear physics and astrophysics. 
 One of the major challenging open problem is what are  the limits of the existence of nuclei.
 A lack of detailed understanding of nucleon-nucleon interaction and its co-relation are 
the  main reasons behind it.
Experimental observations of the failure of validation of 'old magic number' \cite{my49}
 in the nuclei near drip-line further 
indicate that also \cite{Ref a,datta16}. Recently, it has been notified that a number of ingredients in
 nucleon-nucleon interaction such
 as  spin-isospin  monopole interaction,  three body interaction, tensor interaction part \cite{ots,utsuno} etc.
 are essential to
 explain experimentally  observed  properties of nuclei around the  drip-line.  The experimental data on nuclear shell
 structure around drip line  is very essential  and may provide important information on nucleon nucleon interaction. 
Investigation of nuclear shell structure around magic numbers are particularly important in this respect. 
'Island of Inversion' are neutron-rich Ne, Na, Mg nuclei around $N\sim$ 20 where first failure of magic number had been
 reported \cite{Ref a}. It has been observed that the ground state properties of these nuclei 
can be explained only by considering $pf$ shell contribution in addition to $sd$ shell contribution and often, 
it has been concluded that these nuclei exhibit large deformation.  
The state of the art of  Coulomb breakup of the neutron-rich nuclei  has been used 
to explore those properties. The unambiguous information on detailed 
components of the ground-state wave-function along with quantum numbers of the valence 
neutron of the nuclei has been  obtained from the measurement of threshold strength along
 with the $\gamma$-rays spectra of the core following Coulomb breakup \cite{Ref e} .
In this article, several previous successful measurements will be mentioned to explain validity of the method.
New experimental results on the  ``microscopic picture`` of the ground state wave-functions of 'Island of Inversion' nuclei, 
explored through Coulomb breakup  will be presented \cite{datta16,anisur17,santosh17}. 
 Coulomb dissociation (CD) is a successful indirect measurement for capture cross-sections of the neutron-rich nuclei
which may provide input to the model for the stellar evolution process \cite{datta07}.
 Some valuable bulk properties of the neutron-rich nuclei like the density dependent symmetry energy, 
 neutron skin etc. play a key role in understanding cosmic phenomena \cite{armin,klim07,ag12}. 
Electromagnetic excitation using intermediate energy radioactive beam provides a unique opportunity 
to access key information to understand those cosmic phenomena \cite{armin,klim07}.
 A preliminary experimental investigation on the bulk properties of  the neutron-rich nucleus, $^{32}$Mg are reported.
\begin{figure}[h]
\includegraphics[width=6.0cm,clip]{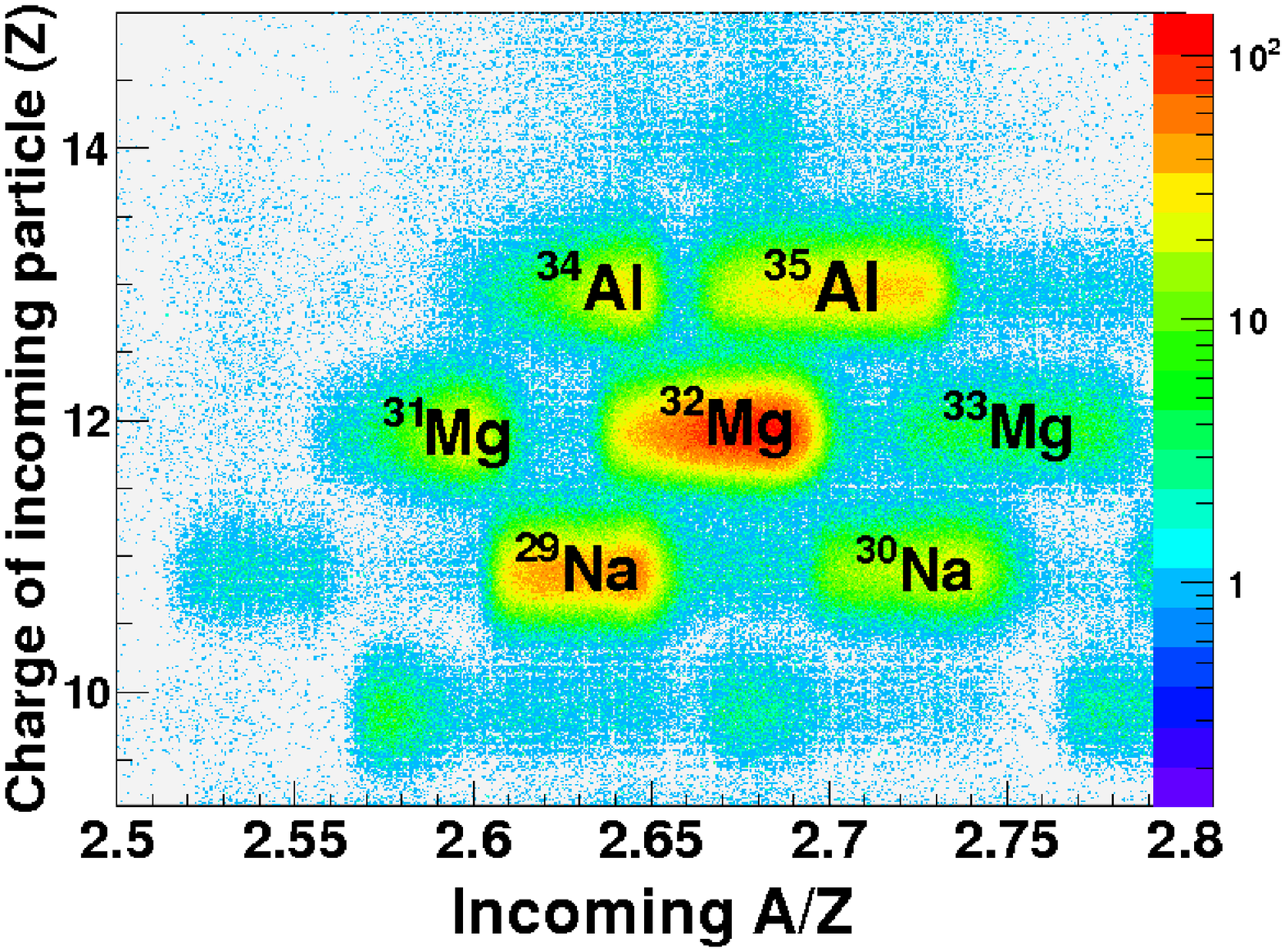}
\includegraphics[width=6.4cm,clip]{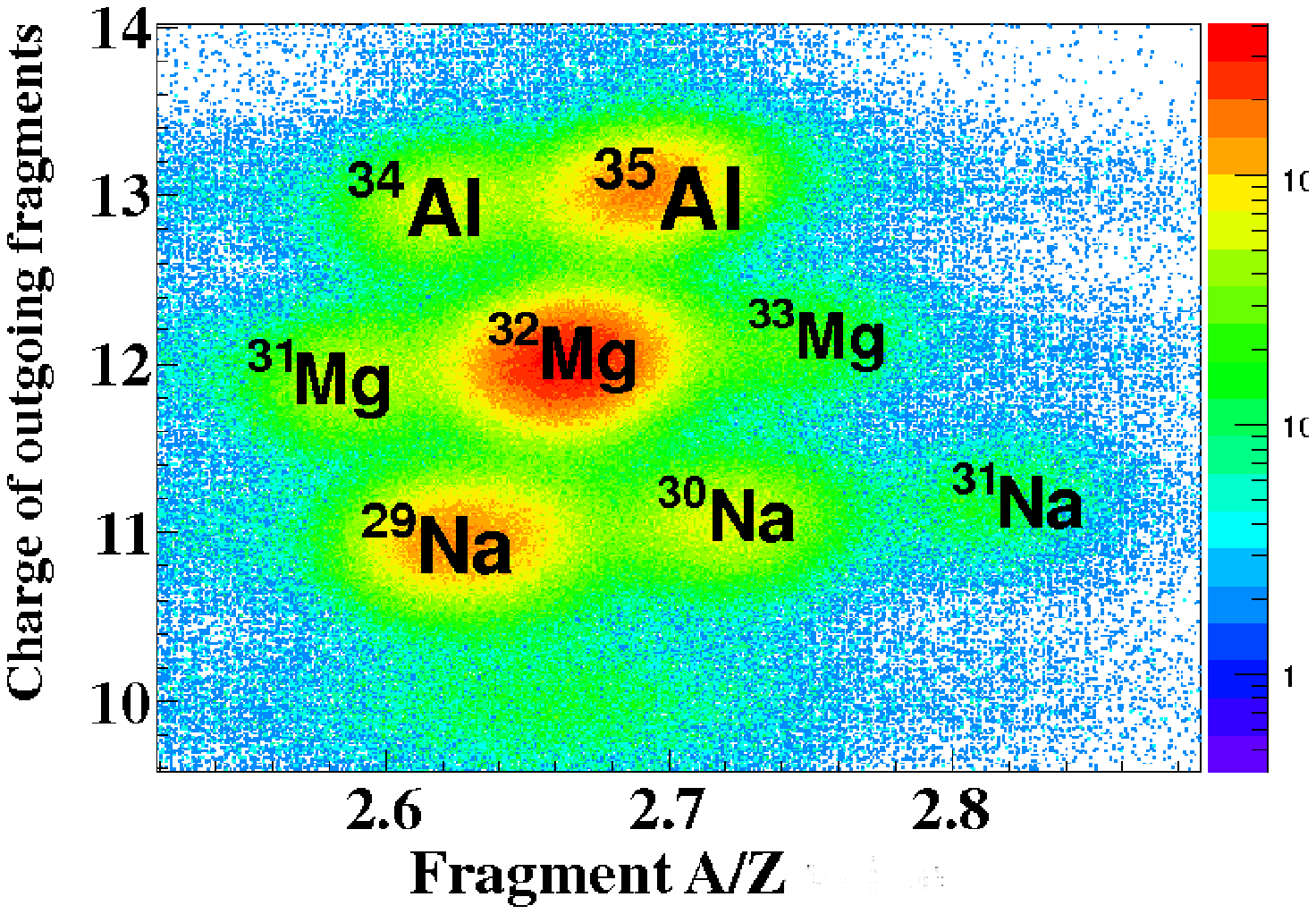}
\caption{Identification plot of cocktail incoming beam [left] and   reconstructed outgoing beam [right]
  of the experiment GSI-S306.}
\vspace{-0.3 cm}
\label{outmass}
\end{figure}
\section {Experimental setup and analysis}
Short-lived radioactive nuclei were produced by the fragmentation of the primary beam  $^{40}$Ar 
at 530 MeV/u   on a $^9$Be production target (8 g/cm$^2$) at GSI, Darmstadt. The secondary beams, 
($^{29-31}$Na, $^{31-33}$Mg, $^{34,35}$Al etc.) with A/Z between 2.55 to 2.85 (Fig. \ref{outmass}-left)
 were separated  according to the magnetic rigidities by the fragment separator (FRS).
 The secondary cocktail beam was bombarded on lead (2 g/cm$^2$) and carbon  (0.9 g/cm$^2$) targets 
 for studying electromagnetic and nuclear   excitation, respectively.  The $\gamma$-rays 
from the excited projectile or fragments were detected by the 4 $\pi$-crystal
 ball spectrometer. Eight double-sided Silicon Strip Trackers (SSTs) were placed enclosing the
reaction target  in 4$\pi$ solid angle. After reaction at the secondary target, reaction fragments  and neutron(s) 
 are forward focused due to Lorentz boost and pass through A Large Dipole Magnet (ALADIN). Reaction fragments, 
deflected by ALADIN according to their A/Z ratios, were tracked by two scintillating fiber 
detectors (GFIs) and detected by the time of flight wall (TFW). The trajectories of neutrons
remain unchanged and were detected by the Large Area Neutron Detector (LAND). Fig. \ref{outmass}-right
shows a reconstructed outgoing mass identification plot  after the secondary target.
Data analysis has been performed using CERN-ROOT platform and programs developed at 
GSI, Darmstadt  and SINP, Kolkata \cite{datta16,anisur17,santosh17}. 
\begin{figure}
\includegraphics[width=5.4cm,clip]{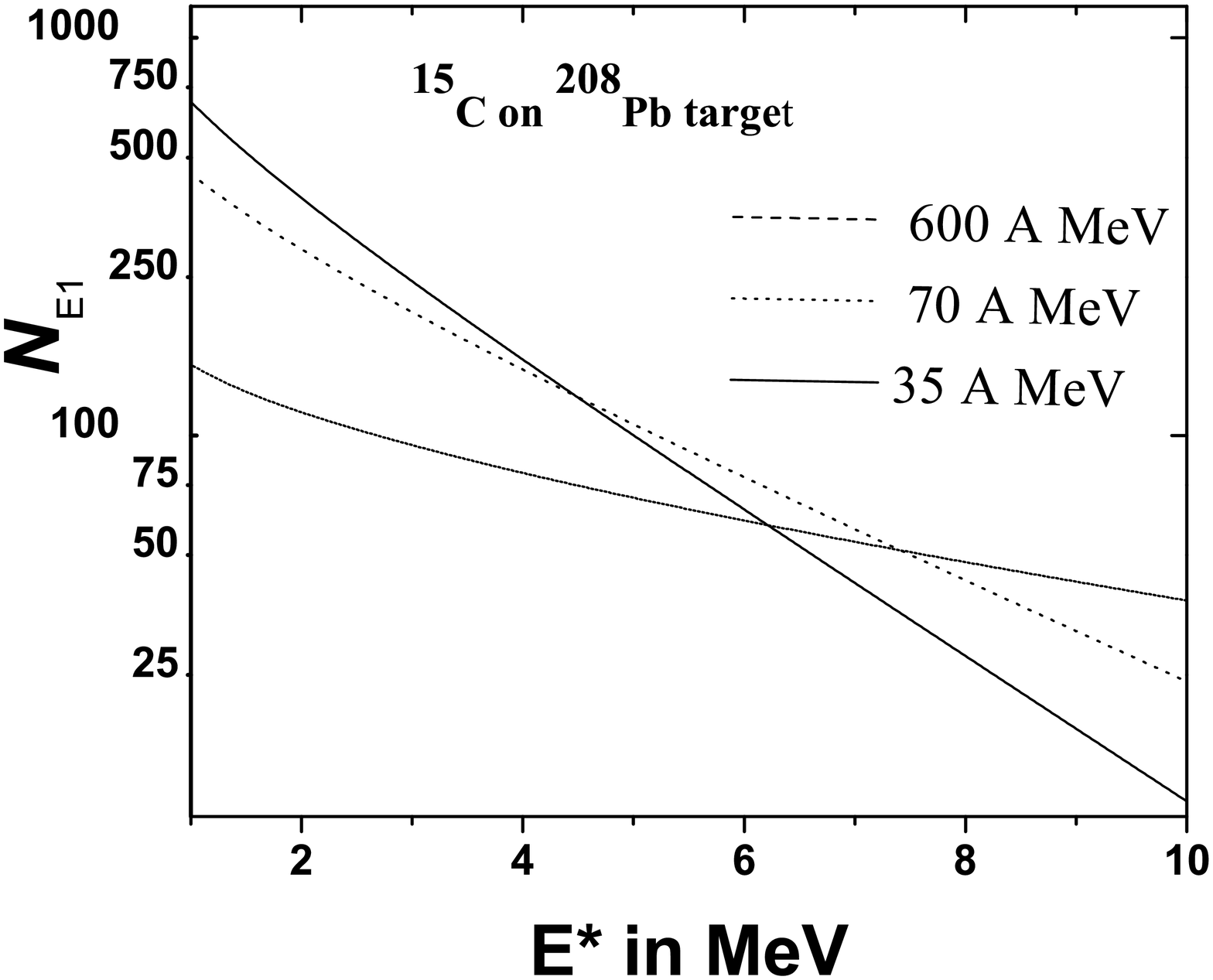}
\includegraphics[width=4.5cm,clip]{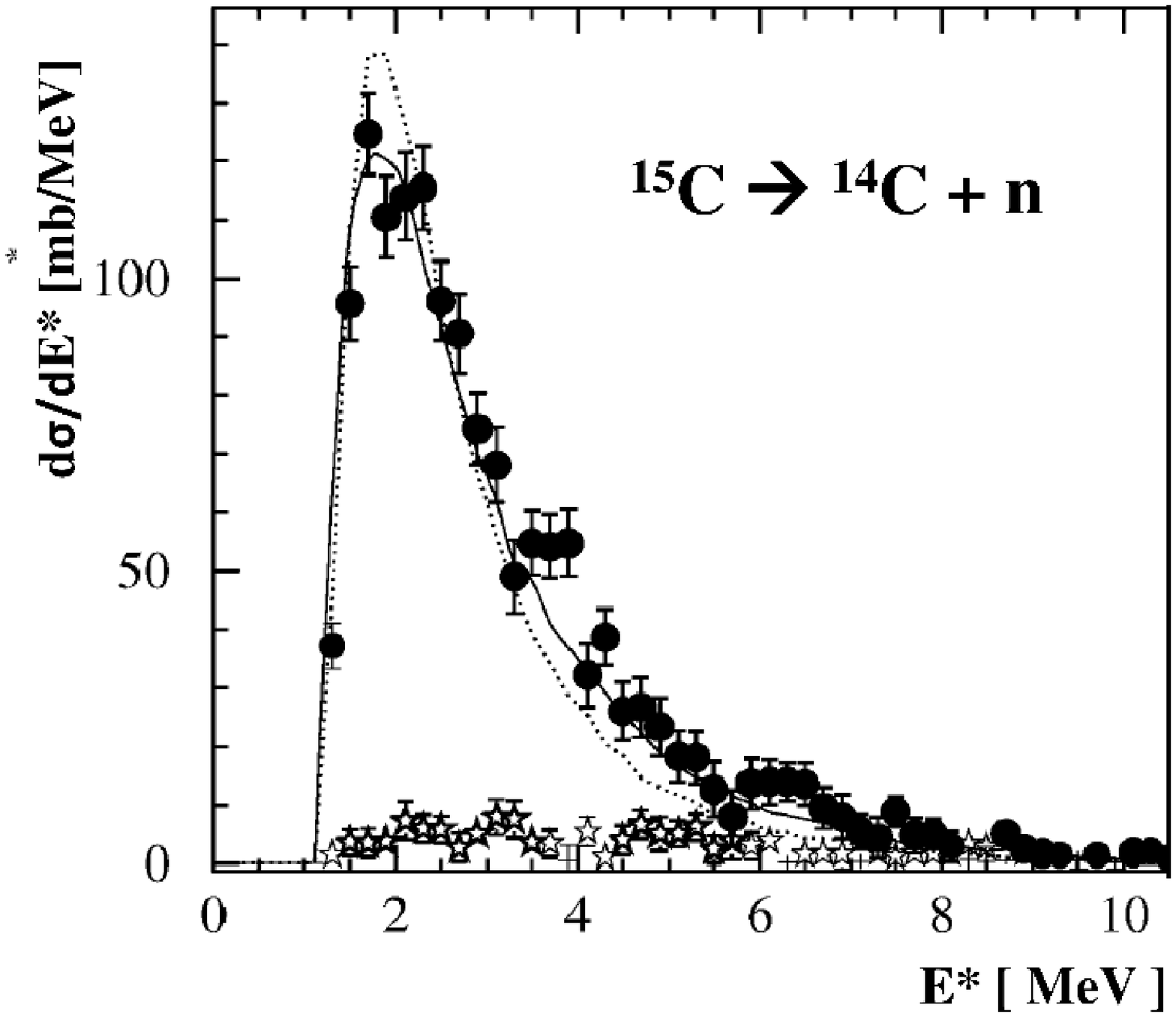}
\includegraphics[width=5.1cm,height= 4.1cm,clip]{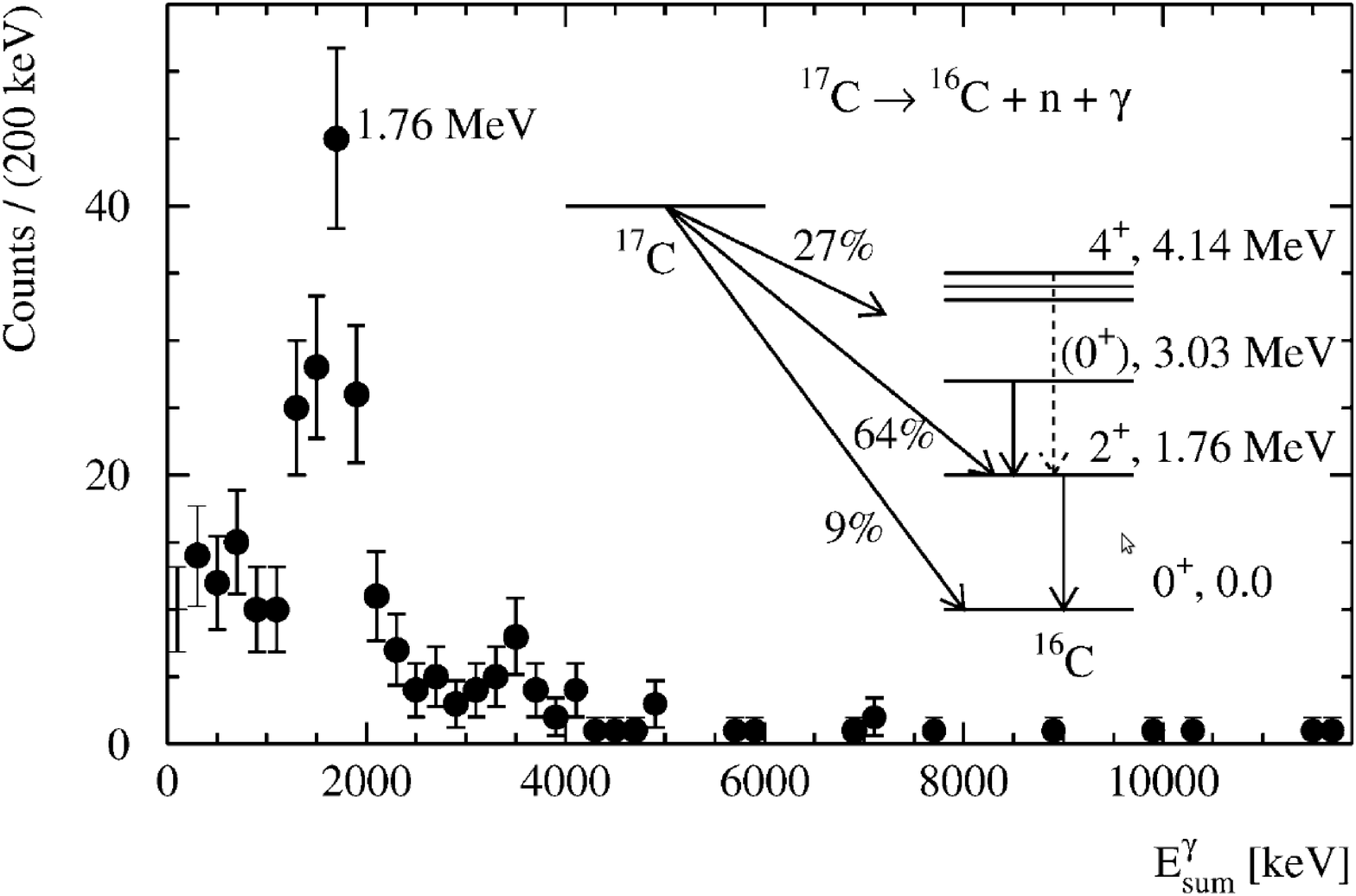}
\caption{[Left] E1 virtual photon spectrum for $^{15}$C projectile of 35 MeV/u, 70 MeV/u and 600 MeV/u
energies respectively, on a Pb target. [middle] Differential Coulomb dissociation (CD)
cross-section with respect to the excitation energy (E $^\star$ ) of $^{15}$C  breaking
up into a neutron and a $^{14}$C fragment in its ground-state (filled
circles). The solid curve displays the result from the direct-breakup
model in a plane-wave approximation with the valence neutron occupying the s-wave. 
[right] The sum-energy spectrum of 
$\gamma$-decay transitions from $^{16}$C after Coulomb breakup of $^{17}$C. 
The inset shows a partial scheme of levels in $^{16}$C and their population after Coulomb breakup of  $^{17}$C.
 Figure [middle and right] reprinted from Datta Pramanik et al. PLB 551, 63 (2003).}
\label{c15}
\end{figure} 
The excitation energies $E^\star$  of  the neutron-rich Na, Mg and Al nuclei, were reconstructed event by event
 by measuring four momenta of all decay products
of those nuclei  after  breakup at the secondary target via invariant mass analysis, The Coulomb
dissociation (CD) cross section for the $^{208}$Pb target (2.0 g/cm$^2$) has been determined after subtracting the 
nuclear contribution which was obtained from the data with a $^{12}$C target (0.9 g/cm$^2$ ) with proper a scaling factor.
Background contributions from reactions induced  by detectors   materials  were determined from data 
taken without any target and were subtracted. The state of the art of one neutron threshold strength via Coulomb dissociation 
has been employed to explore single particle properties of the neutron-rich nuclei.  
On the other hand electromagnetic excitation via inverse kinematics 
of the neutron-rich nuclei up to several   neutrons and proton threshold  energy provide access to several bulk properties 
of neutron-rich nuclear matter. Below we are discussing in separate sub-section both single particle and bulk 
properties of the neutron-rich nuclei. We shall discuss impact of those properties on cosmic 
phenomena in another sub-section.
\section {Threshold strength and single-particle properties of neutron-rich nuclei}
When a projectile  with relativistic energy passes a high Z target, it  may be excited by absorbing virtual photons
from the time dependent Lorentz-contracted Coulomb field \cite{ber88} and breakup into a core and a neutron. 
 This one neutron removal differential CD cross-section can be expressed by
the following equation \cite{Ref e}:
\vspace{-0.5 cm}
\begin{equation}
\frac{d\sigma_c}{dE^*}=\ \frac{16\pi^3}{9\hbar c}N_{E1}(E^*)\displaystyle\Sigma_j C^2S(I^{\pi}_c,nlj)
  \\\times \displaystyle\Sigma_m|<q|(Ze/A)rY^l_m|\psi_{nlj}(r)>|^2 \
\end{equation}
 $\psi_{nlj}(r)$ and $<q|$ represent the single-particle
 wave function of the valence neutron in  the projectile ground state and the final state in the continuum 
respectively.
\begin{wrapfigure}{r}{7.0cm}
\includegraphics[width=7.0cm ]{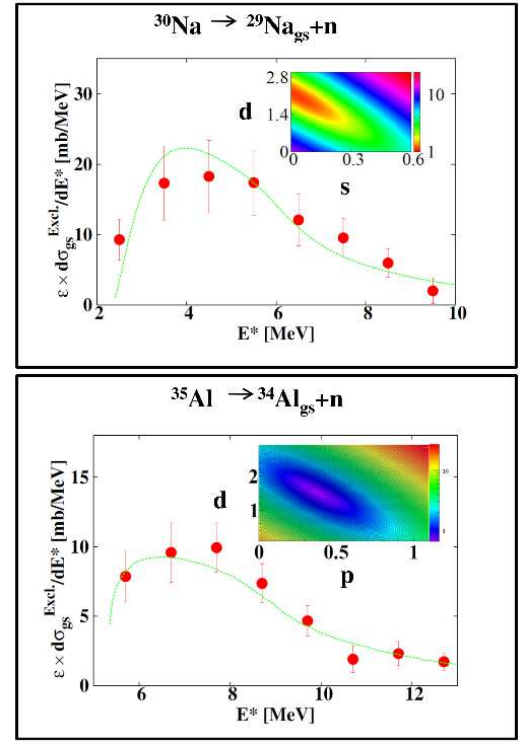}
\caption{ [top] The differential CD cross-section with respect to
 the excitation energy (E$^\star$) of $^{29}$Na [top] and of $^{35}$Al [bottom]
 breaking up into a neutron and a fragment in its ground state or isomeric state  
(filled circles). The solid curve displays the result from the direct-breakup
model where the valence neutron(s) is occupying $s,d$  
and $p,d$ orbitals  for the top-panel and the bottom-panel of the figure, respectively.}
\label{inv29}
\end{wrapfigure}
 $C^2S(I^{\pi}_c,nlj)$ represents the spectroscopic factor
 with respect to a particular core state $I^{\pi}_c$ of that. 
$N_{E1}(E^\star)$ is the number of virtual photon as a function of excitation energy E$^\star$  \cite{ber88}. 
Fig. \ref{c15}-left  shows the  variation of the
E1 virtual photon spectrum  with different 
 projectile energies for $^{15}$C on Pb target. It is evident from  the virtual photon spectra that
 the number of electric dipole virtual photons, $N_{E1}$ decreases with excitation energy (E$^\star$) for low energy projectile 
due to the adiabatic cutoff. 
Fig. \ref{c15}-middle  shows the measured  differential CD cross-section at an energy 600 A MeV with respect 
to excitation energy (E $^\star$) of  $^{15}$C breaking up
 into a neutron and a $^{14}$C fragment in its ground state (filled circles) \cite{Ref e}. 
The solid curve displays the cross-section obtained from the direct-breakup (DB) [equation (1)]
model considering the valence neutron occupying the s$_{1/2}$ orbital. 
Thus it is evident from the Fig. \ref{c15} that the data is in agreement with the configuration 
$^{14}$C$_{gs}\otimes\nu_{s1/2}$
and the spectroscopic factor of the valence neutron orbital is in agreement with the  transfer reaction and knockout
reaction data (see \cite{Ref e} for details). However, in some of the neutron-rich nuclei due to specific effects
of the n-n interaction, various types of particle-hole excitation are possible in the ground state of neutron-rich nuclei.
Those components can be identified by measuring $\gamma$-rays in coincidence with core after CD. That was demonstrated
by measuring CD of $^{17}$C. Fig. \ref{c15}-[right] shows the sum-energy spectrum of 
$\gamma$-decay transitions from $^{16}$C after Coulomb breakup of $^{17}$C. The inset shows a partial scheme of
levels in $^{16}$C and their population after Coulomb breakup.  The ground state spin and parity of exotic nuclei
can be obtained by coupling the spin and parity of the core state  with that of valence neutron.
\subsection {Shell evolution in the  neutron-rich nuclei}
\begin{figure}[h!]
\includegraphics[width=7.5cm,height=5.5cm ]{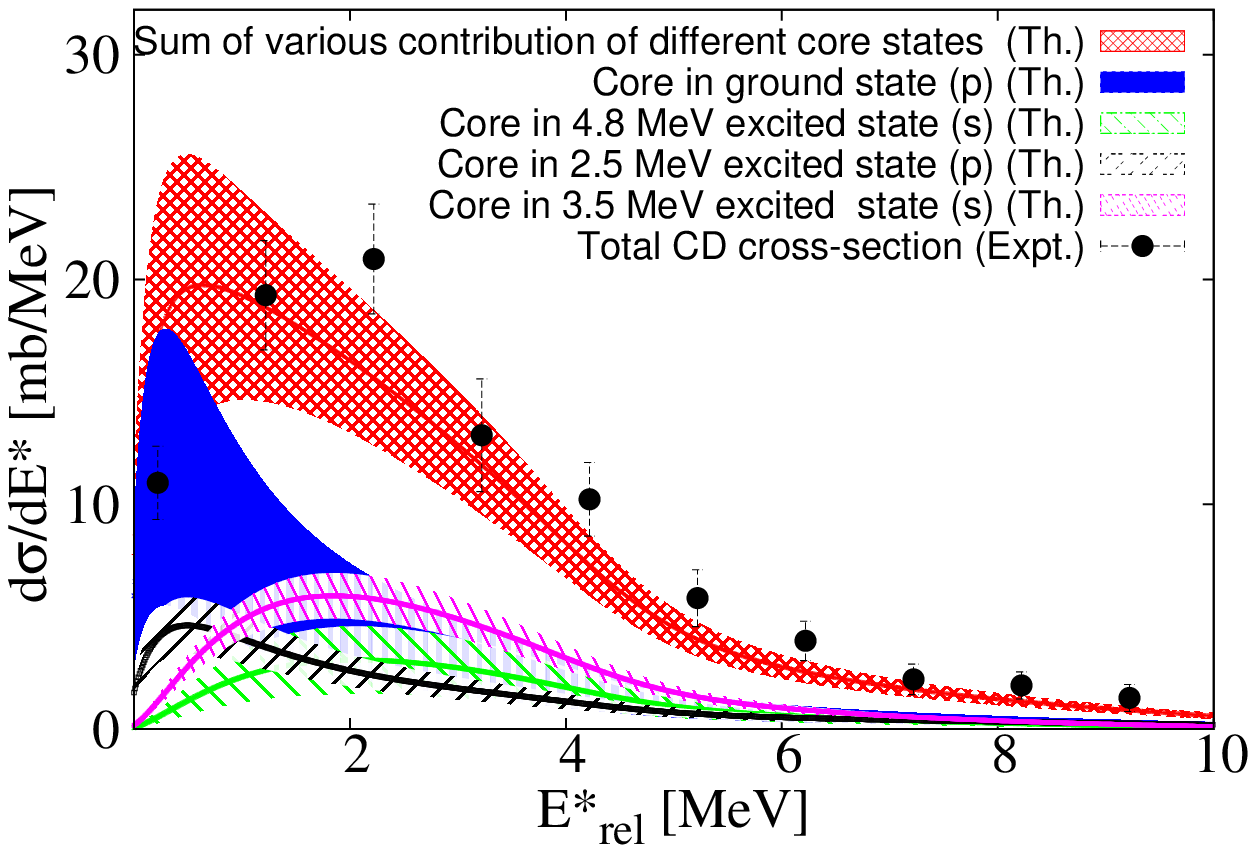}
\includegraphics[width=8.0cm]{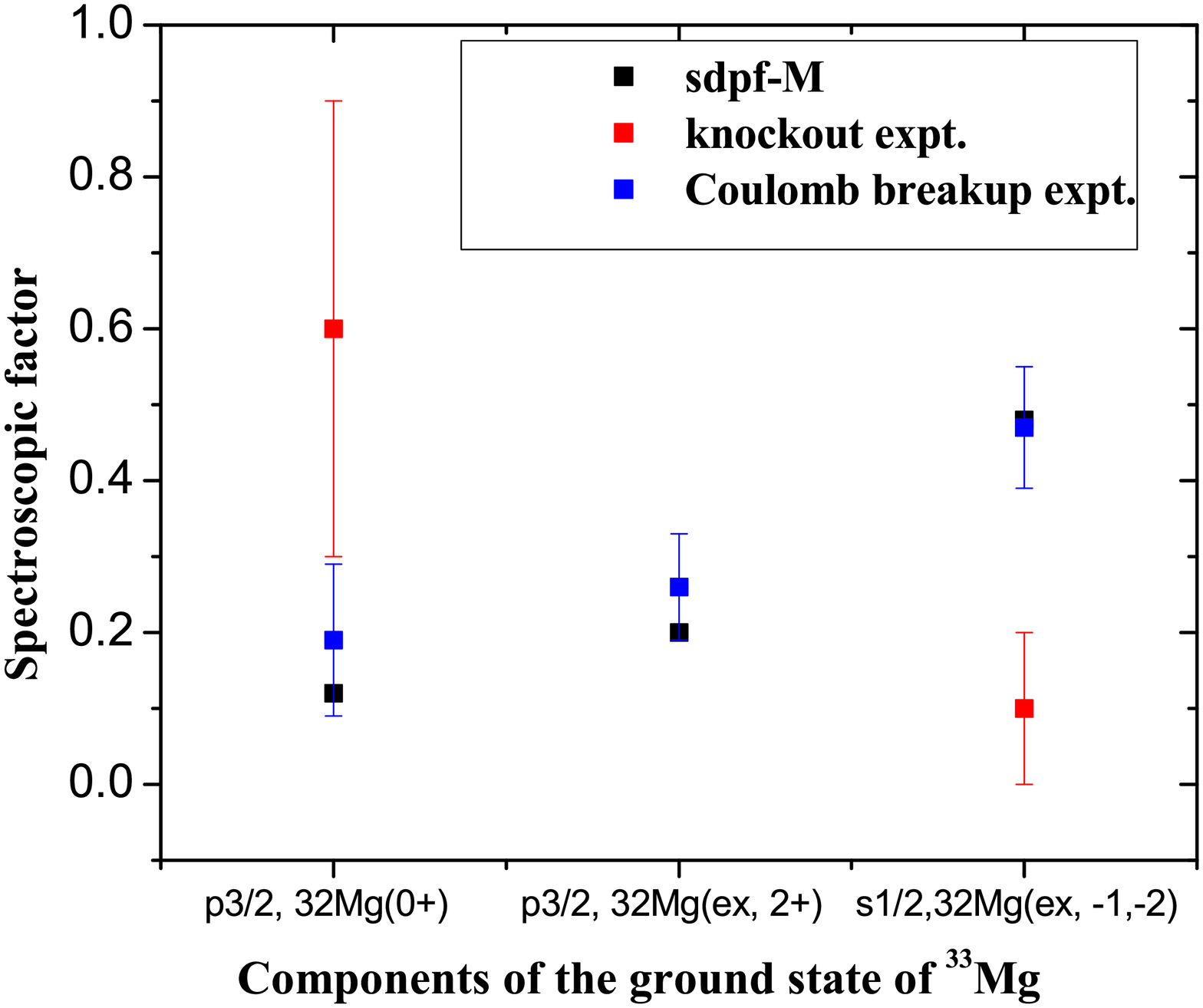}
\vspace{-0.3 cm}
\caption{ [left] The total experimental inclusive (black filled-circle data)
differential CD cross-section of $^{33}$Mg is overlaid with the sum
of calculated exclusive CD cross-sections (red solid line) for four
components of the ground-state. The (red on-line) mesh shaded region
is the associated errors obtained from the various contributions.
Other shaded regions as mentioned in the figure represent different
components of the  ground state wave function obtained from exclusive 
CD cross-sections with the core excited states. Figure reprinted from 
Ushasi Datta et al, PRC 94, 034304 (2016). [right] The spectroscopic factors
for  the  occupying orbitals obtained from $spdf$-M calculation [black square], 
knockout reaction \cite{kan10} [red square] and 
Coulomb breakup measurement [blue square] \cite{datta16} are shown.}
\label{33mg}
\end{figure}

The modification in the shell gaps through  effects such as the tensor component of the 
NN force become pronounced with large neutron-proton asymmetries in the exotic nuclei 
far away from stability. These lead to the disappearance of established magic numbers and 
the appearance of new ones. The first observation of the disappearance of so-called magic 
number ($N$ = $20$) was reported, based on the mass measurements in
neutron-rich $^{31,32}$Na   and BE2 measurement in $^{32}$Mg \cite{Ref a}. For a deeper understanding
of the n-n interaction  microscopic information on the ground state wave-function is essential.
The state of the art of Coulomb breakup has been utilized to explore the detailed components of 
the ground-state wave-function of the neutron-rich Na, Mg, Al around N$\sim$ 20. 
  Monte-Carlo shell  model calculation \cite{ots} has predicted  the lowering
 of 2p$_{3/2}$ orbital for aluminum  isotopes around the N$\sim$20 shell gap. On the other hand to explain 
the knockout data \cite{kan10,utsuno}
for $^{33}$Mg, the shell gap was reduced from 6 MeV to 3 MeV in $sdpf-M$ calculation. We  describe 
below our observation in this respect via Coulomb breakup method.\\
{\bf $^{29,30,31}$Na:}\\
The first results were reported  on the ground state configurations of the neutron-rich $^{29,30}$Na isotopes, obtained via  
CD  measurements \cite{anisur17}.  
A comparison  with the direct breakup model,  suggests the predominant components of the  ground state configurations of 
  these nuclei are   $^{28}$Na$_{gs}(1^+)\otimes\nu_{d}$ and 
$^{29}$Na$_{gs}(3/2^+)\otimes\nu_{d}$, respectively. The ground state spin and  parity of these nuclei obtained from this 
experiment are in agreement with earlier reported values. The spectroscopic factors for the valence neutron occupying 
the $d$ orbital for these nuclei in the ground state have been  extracted and reported for the first time.
This is  in agreement with a USDB calculation for $^{29}$Na but that the  factor is reduced in comparison with
the  USDB for $^{30}$Na. 
 A comparison of the experimental  findings with  shell model calculation using the  MCSM suggests a lower limit of 
around $4.3$ MeV of the $sd-pf$ shell
 gap in $^{30}$Na.  In the case of $^{31}$Na, the major component of the ground state configuration ($\sim$ 60 \%) is  multi-particle-hole in nature.
which  indicates a  narrower $N =$ 20 shell gap. \\
{\bf $^{31,33}$Mg}:\\
The valence neutron in the ground state of $^{33}$Mg should occupy the $f_{7/2}$ orbital as in  $^{41}$Ca. 
But experimental evidence
 is different. We reported the first direct experimental evidence  of a  multi-particle-hole  ground-state 
configuration obtained  via a (400 A MeV) CD measurement \cite{datta16}.   
The major part  $\sim (70\pm13)\% $ 
 of the cross-section is observed to populate  the excited states of  $^{32}$Mg after the CD. 
Various components of the wave-function, obtained from this experiment are shown in Fig.\ref{33mg}.
The shapes of the differential CD cross sections in coincidence   
with  different  core excited states   favor that the  valence neutron  occupies both  the $s_{1/2}$  and $p_{3/2}$ orbitals. 
These experimental findings suggest a significant reduction and merging of  the  $sd-pf$ shell  gaps at $N$  $\sim$ 20 and 28.  
 The experimentally obtained quantitative spectroscopic information for the valence neutron occupation of  
the  $s$ and $p$ orbitals, coupled with different core states
 is in agreement with MCSM calculation using 3 MeV as  the shell gap at $N$ = 20 (Fig.\ref{33mg}-right).
\vspace{-0.5 cm}
\\
{\bf $^{34,35}$Al}:\\
Similarly, the valence neutron in $^{35}$Al ($N$ =22) should occupy the $f_{7/2}$ orbital but 
   the observed  differential CD cross section of $^{35}$Al $\rightarrow ^{34}$Al + n could not  be
interpreted in the light of direct breakup model  with that configuration. It suggests that 
 the  possible ground state 
spin and parity of $^{35}$Al could be tentatively,  1/2$^+$ or  3/2$^+$ or 5/2$^+$.  
If  5/2$^+$ is the ground state spin/parity of $^{35}$Al 
as suggested in the literature, then the major ground state configuration of $^{35}$Al is a combination of
 $^{34}$Al$(g.s.;4^-)\otimes \nu_{p_{3/2}}$ and $^{34}$Al$(isomer; 1^+)\otimes \nu_{d_{3/2}}$ states.
The result from this experiment has been compared with that from previous knockout measurement and 
$sdpf$-M calculations \cite{santosh17}. This hints at a  particle-hole configuration of the  
neutron across the magic shell gaps  at $N$ = 20,28 
which suggests narrowing the magic shell gap. Fig. \ref{inv29} shows  the differential CD cross section with 
respect to excitation energy (E $^\star$ ) of $^{30}$Na  [top]  and  $^{35}$Al [bottom]
 breaking into core  and  neutron. The solid line 
represents the calculation where valence neutron is occupying  the $s$ and $d$-wave for $^{30}$Na and 
the $p$ and $d$-wave for $^{35}$Al.  
\section{Bulk properties of neutron-rich nuclei}
\begin{figure}[t!]
\includegraphics[width=12cm,height=4.6 cm ]{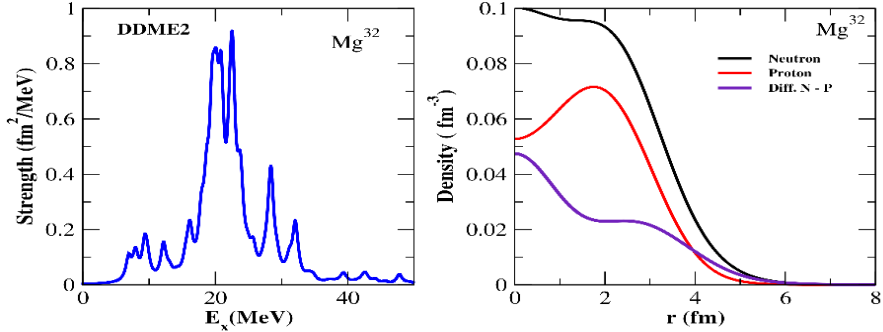}
\caption{[left]The strength function for the isovector giant resonance  in $^{32}$Mg 
obtained from a density dependent meson exchange model, DDME2 \cite{nil07}. 
[right] The density profile for  protons, neutrons and the difference between neutrons and protons,
obtained from the calculation are shown.}
\label{32mg_th}
\end{figure}
\begin{wrapfigure}{r}{7cm}
\includegraphics[width=7cm ]{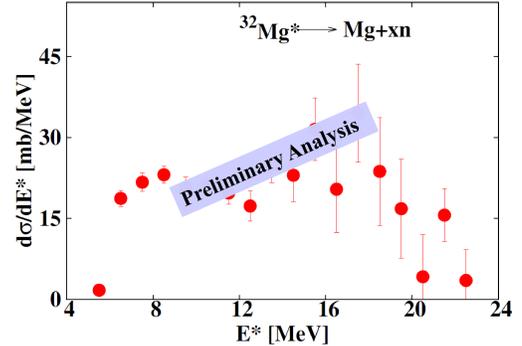}
\caption{ Differential CD cross section with respect to
 E$^\star$ of $^{32}$Mg breaking
up into  neutron(s) and ($^{30,31}$Mg) fragment.}
\label{32mg_expt}
\end{wrapfigure}

 
The bulk properties of neutron-rich nuclei would access valuable
information on n-n co-relations, the validation of new mean-field calculations
 with different interactions,
the density dependent symmetry energy etc. \cite{armin,klim07,ag12,nil07}.
Relativistic mean-field calculations using a
Density Dependent Meson Exchange coupling  (DDME2) \cite{nil07} have been used to calculate
the strength function for the isovector giant dipole resonance of $^{32}$Mg as a function
of excitation energy. In this model, only the linear terms for the mesonic
fields are considered. The non linear contribution of the meson fields
are realized  through the density dependence of the coupling constants
for the nucleon-meson interactions. Fig.\ref{32mg_th} shows the IVGDR response
and density profile  for $^{32}$Mg nucleus, calculated by DDME2. This calculation predicts 
a neutron skin of 0.3 fm. As predicted theoretically,  
additional low-lying dipole strength  
below the giant resonance in a number of  neutron-rich nuclei has been observed \cite{armin}.  
This strength  is known as the pigmy resonance .
The  excitation energy of  $^{32}$Mg has been studied via electromagnetic excitation using 
intermediate energy RIB. 
Preliminary data analysis shows alow-lying dipole or pigmy resonance at an 
excitation energy around 7-11 MeV 
(Fig. \ref{32mg_expt}) which may provide information on neutron skin thickness.
 A detailed comparison between 
the data and new mean-field calculations with a modified interaction may provide insight 
of n-n interaction
with large isospin degrees of freedom. A more details analysis is in progress. 
\section{Neutron-rich nuclei and cosmic phenomena}
The nuclei around the drip-line are short-lived but surprisingly, 
evanescent rare isotopes imprint their existence in stellar explosive scenarios (r-process etc.).
 Due to their fleeting existence, indirect measurements are often the only possible access 
to the information which is a valuable input to the model for star evolution process.
A number of indirect methods are being explored by experimental nuclear physicists to avoid 
radioactive targets and other difficulties of
direct measurements of radiative capture cross sections \cite{datta07}.
 The Coulomb dissociation of radioactive ion beams
at intermediate energy is one of the most powerful indirect methods for measuring capture cross sections,
and is being explored at various laboratories in the world. $^{14}$C(n,g)$^{15}$C  was studied by direct and indirect 
method and a detailed comparison with both methods along with various theoretical calculation was performed and results
are consistent \cite{datta07}. 
The recent discovery of  neutron stars merging \cite{ligo}  provides  a center of attention among
nuclear physicists to study the equation of state of neutron-rich matter.
To understand the properties of   neutron stars, neutron star merging etc., information on the neutron skin 
and the density dependent symmetry energy at saturation density is important\cite{ag12,nil07}.
 A number of experimental approaches
are being pursued \cite{klim07,betty,prex}. By measuring low lying dipole strength and dipole-polarizability,
 neutron skin measurements have been attempted \cite{klim07}.
But a better co-relation  among physical quantities is still being  searched for. We are investigating in this respect 
to further improve information on neutron skin from the  $^{32}$Mg data.
\section {Conclusion} 
The state of the  art of Coulomb breakup has been used to explore both single particle and bulk properties of the neutron-rich
nuclei Na, Mg, Al  around the 'island of inversion' where the first disappearance of magic number
 has been observed. 
The one neutron dipole  threshold strength  is sensitive to the
quantum numbers and binding energy of  the valence neutron. Using that property, 
clear evidence has been observed for the  melting and merging of the magic
 shell gaps at $N $= 20, 28 in $^{31}$Na, $^{33}$Mg and $^{35}$Al. On the other 
hand the predominant ground-state 
configurations of $^{29,30}$Na  redefines the boundary of 'Island of Inversion'.  
''r-process" has been known as a basic formation process for the
heaviest elements .  CD can be used for  indirect measurements  of the capture cross-sections
of the neutron-rich nuclei  which may provide valuable information to  model the r-process.
 Some valuable bulk properties of the neutron-rich nuclei like the density dependent symmetry energy, 
 the neutron skin etc. play a key role in understanding cosmic phenomena.  The  excitation energy of  
$^{32}$Mg has been studied. A preliminary data analysis shows low-lying dipole or Pigmy 
resonance at an excitation energy around 7-11 MeV which may provide information on its neutron 
skin thickness. Details analysis are in progress. 

\end{document}